\documentclass{article}
\usepackage{graphicx}
\usepackage{color}
\usepackage{amsmath, amsthm}
\usepackage{amssymb}
\usepackage{rotating}
\usepackage{epsfig}
\usepackage{a4}
\input amssym.def
\input amssym.tex
\newtheorem{theorem}{Theorem}[section]
\newtheorem{corollary}{Corollary}[section]
\newtheorem{proposition}{Proposition}[section]

\def\ben{\begin{equation}}
\def\een{\end{equation}}

\def\bea{\begin{eqnarray}}
\def\eea{\end{eqnarray}}

\title{On the nonlinear stability of higher-dimensional triaxial Bianchi IX black holes}
\author{Mihalis Dafermos\thanks{University of Cambridge,
Department of Pure Mathematics
and Mathematical Statistics, Wilberforce Road, Cambridge CB3 0WB, United 
Kingdom}
\and Gustav Holzegel\thanks{University of Cambridge,
Department of Applied Mathematics
and Theoretical Physics, Wilberforce Road, Cambridge CB3 0WA, United 
Kingdom}}
\begin{document}
\maketitle
\begin{abstract}
In this paper, we prove that the $5$-dimensional Schwarzschild-Tangherlini
solution of the Einstein vacuum equations
is orbitally stable (in the fully non-linear theory) with respect to 
vacuum perturbations
of initial data preserving triaxial Bianchi-IX symmetry. 
More generally, we prove that $5$-dimensional vacuum spacetimes developing
from suitable asymptotically flat triaxial Bianchi IX symmetric initial data 
and containing a trapped or marginally trapped homogeneous $3$-surface necessarily
possess a complete null infinity $\mathcal{I}^+$, whose past $J^-(\mathcal{I}^+)$
is bounded to the future by a regular event horizon 
$\mathcal{H}^+$, whose cross-sectional volume in turn satisfies a Penrose
inequality, relating it to the final Bondi mass.
In particular, the results of this paper give
the first examples of vacuum black holes
which are not stationary exact solutions. 
\end{abstract}

\section{Introduction}
The study of higher-dimensional gravity has 
attracted much attention in recent years, motivated 
mainly by speculations from high energy physics.
The variety of possible end-states for vacuum gravitational 
collapse in higher dimensions appears richer~\cite{Harvey} than in $4$ dimensions and 
gives rise to many interesting questions. 
All analytical work, thus far, however, has centred on the question of
the existence and uniqueness of static~\cite{Daisuke} or 
stationary~\cite{MyersPerry, gang} solutions, or has been based on
study of the linearized equations~\cite{Hartnoll, Akihiro}. 
While such results are suggestive 
as to what may occur dynamically, they do not directly address
the problem of evolution and leave open the possibility that 
the non-linear theory admits phenomena of a completely different and 
unexpected nature.

The purpose of this paper is to initiate the rigorous study 
of dynamical 
vacuum black holes in higher dimensions in the \emph{fully non-linear} theory.
Specifically, we will study the problem of evolution for the Einstein
vacuum equations
\begin{equation}
\label{vacuum}
R_{\mu\nu}=0,
\end{equation}
 for asymptotically flat initial data possessing
triaxial Bianchi IX symmetry. This model has been
recently introduced by Bizon et.~al.~\cite{Bizon}.
They show that vacuum solutions 
with this symmetry have
two dynamic degrees of freedom, and the Einstein equations can be 
written (see~\cite{Bizon})
as a system of non-linear pde's on a $2$-dimensional Lorentzian quotient
of $5$-dimensional spacetime by an $SU(2)$ action with
$3$-dimensional orbits.

The system of equations thus obtained is
studied numerically in~\cite{Bizon}, where analogues
of critical behaviour have been discovered. 
Proving rigorously the kind of behaviour suggested by these numerics
appears a formidable problem, beyond the scope of current techniques. 
Implicit in the discussion of~\cite{Bizon}, however, is the notion that there
is an open set of initial data that leads to black hole formation.
It is this aspect of~\cite{Bizon} that we will formulate
and rigorously prove in this paper.

The main result is
\begin{theorem}
\label{introthe}
Consider asymptotically flat smooth initial data $(\mathcal{S},\bar{g},K)$ 
for the vacuum Einstein equations,
possessing triaxial Bianchi IX symmetry. Let $(\mathcal{M},g)$ denote the maximal
Cauchy development, and let $\pi:\mathcal{M}\to \mathcal{Q}$ denote 
the projection map to the $2$-dimensional Lorentzian quotient $\mathcal{Q}$.
Suppose there exists an asymptotically flat
spacelike Cauchy surface $\tilde{S}\subset\mathcal{Q}$,
and a point $p\in\tilde{S}$ such that $\pi^{-1}(p)$ is trapped or marginally
trapped, and (at least) one of the connected components
$\tilde{S}\setminus\{p\}$ containing an asymptotically flat end is such that
$\pi^{-1}(q)$ is not outer antitrapped or marginally antitrapped 
for any $q$ in the component.
Then $\mathcal{Q}$
contains a subset with Penrose diagram:
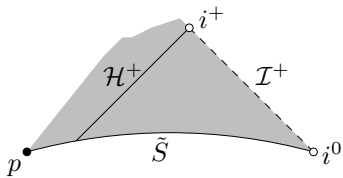
\[
\begin{picture}(0,0)%
\includegraphics{geg.pstex}%
\end{picture}%
\setlength{\unitlength}{3158sp}%
\begingroup\makeatletter\ifx\SetFigFont\undefined%
\gdef\SetFigFont#1#2#3#4#5{%
  \reset@font\fontsize{#1}{#2pt}%
  \fontfamily{#3}\fontseries{#4}\fontshape{#5}%
  \selectfont}%
\fi\endgroup%
\begin{picture}(2467,1315)(2401,-4094)
\put(3526,-3961){\makebox(0,0)[lb]{\smash{\SetFigFont{10}{12.0}{\rmdefault}{\mddefault}{\updefault}{\color[rgb]{0,0,0}$\tilde{S}$}%
}}}
\put(3901,-2911){\makebox(0,0)[lb]{\smash{\SetFigFont{10}{12.0}{\rmdefault}{\mddefault}{\updefault}{\color[rgb]{0,0,0}$i^+$}%
}}}
\put(4868,-3980){\makebox(0,0)[lb]{\smash{\SetFigFont{10}{12.0}{\rmdefault}{\mddefault}{\updefault}{\color[rgb]{0,0,0}$i^0$}%
}}}
\put(3151,-3361){\makebox(0,0)[lb]{\smash{\SetFigFont{10}{12.0}{\rmdefault}{\mddefault}{\updefault}{\color[rgb]{0,0,0}$\mathcal{H}^+$}%
}}}
\put(2401,-4036){\makebox(0,0)[lb]{\smash{\SetFigFont{10}{12.0}{\rmdefault}{\mddefault}{\updefault}{\color[rgb]{0,0,0}$p$}%
}}}
\put(4351,-3361){\makebox(0,0)[lb]{\smash{\SetFigFont{10}{12.0}{\rmdefault}{\mddefault}{\updefault}{\color[rgb]{0,0,0}$\mathcal{I}^+$}%
}}}
\end{picture}

\]
Moreover, the null infinity $\mathcal{I}^+$ corresponding to the above
end is complete, and 
the Penrose inequality
\[
r\le\sqrt{2M_f}
\]
holds on $\mathcal{H}^+$, where $r$ denotes the volume-radius function, and
where $M_f$ denotes the final Bondi mass.
\end{theorem}
\noindent
Note that one can construct a large family of initial data such that
the assumptions of the theorem are satisfied with $\tilde{S}=\pi(\mathcal{S})$,
for instance.

The region $J^-(\mathcal{I}^+)$ 
depicted above is what is typically called a black hole \emph{exterior},
the region $\mathcal{Q}\setminus J^-(\mathcal{I}^+)$ is called the
\emph{black hole}, and $\mathcal{H}^+$ is the \emph{event horizon}.
Thus, the statement of the theorem can be paraphrased by
\vskip1pc
\noindent
\emph{Asymptotically flat triaxial 
Bianchi IX-symmetric spacetimes evolving from suitable data, with
an $SU(2)$-invariant
trapped or marginally trapped
$3$-surface, possess a black hole with a regular event horizon (satisfying
a Penrose inequality) and a complete
null infinity.}
\vskip1pc
\noindent

Theorem \ref{introthe} can in fact be specialized to yield
\begin{corollary}
\label{stab}
Let $(\mathcal{S},\bar{g},K)$ denote initial data evolving to 
the Schwarzschild-Tangherlini metric. Then for smooth triaxial Bianchi IX-symmetric
initial data $(\mathcal{S}',\bar{g}',K')$, sufficiently close to
$(\mathcal{S},\bar{g},K)$ in a suitable norm, the result of the 
previous theorem holds for the maximal development
$(\mathcal{M}',g')$, and moreover, the 
black hole exterior of $(\mathcal{M}',g')$
remains close in a suitable sense to Schwarzschild-Tangherlini.
\end{corollary}

Corollary~\ref{stab} can be paraphrased by the statement:
\vskip1pc
\noindent
\emph{Schwarzschild-Tangherlini is orbitally stable within the class of
triaxial Bianchi IX-symmetric spacetimes.}
\vskip1pc
\noindent
The results of this paper can be thought to
complement previous results of Gibbons and Hartnoll~\cite{Hartnoll} suggesting linear
stability\footnote{See also Ishibashi and Kodama~\cite{Akihiro}.}, 
 and also to the uniqueness of Schwarzschild-Tangherlini
as a static black hole vacuum spacetime~\cite{Daisuke}.
Finally, we note that Theorem~\ref{introthe} gives in particular the first examples
of vacuum black holes which are not static or stationary exact 
solutions.\footnote{Such solutions are yet to be constructed
in $3+1$-dimensions, as, in view of Birkhoff's theorem,
it is impossible to reduce the problem to a $1+1$-dimensional
system of pde's. Solutions with a future complete, but not past complete,
$\mathcal{I}^+$ have been constructed, however, by Chru\'sciel~\cite{Chrusciel},
by solving a certain parabolic problem.}

\section{Triaxial Bianchi IX}
We will say that a globally hyperbolic
spacetime $(\mathcal{M},g)$ admits triaxial Bianchi IX symmetry
if topologically, $\mathcal{M}=\mathcal{Q}\times SU(2)$, for $\mathcal{Q}$ a
$2$-dimensional manifold possibly with boundary, and
where global coordinates $u$ and $v$ can be chosen on $\mathcal{Q}$ such that
\begin{equation} \label{TB}
g= -\Omega^2 du \ dv + \frac{1}{4} r^2
\left(e^{2B} \sigma_1^2 + e^{2C} \sigma_2^2+e^{-2(B+C)} \sigma_3^2\right)
\end{equation}
where $B$, $C$, $\Omega$, and $r$ are functions $\mathcal{Q}\to\mathbb R$, 
and the $\sigma_i$ are a standard basis of
 left invariant one-forms on $SU(2)$, i.e.~such that
coordinates $(\theta,\phi,\psi)$ can be chosen on $SU(2)$ with
\begin{align}
\begin{split}
\sigma_1 &= \sin \theta \sin \psi d\phi + \cos \psi d\theta, \\
\sigma_2 &= \sin \theta \cos \psi d\phi - \sin \psi d\theta, \\ 
\sigma_3 &= \cos \theta d\phi + d\psi.
\end{split}
\end{align}
If there is a boundary $\Gamma$ to $\mathcal{Q}$, it is to be a timelike
curve, characterized by $r=0$.

From the above, it is clear
that the metric (\ref{TB}) admits an $SU(2)$ action by isometry. 
The boundary $\Gamma$ corresponds to fixed points of the group action.
We call it the \emph{centre}.
The angular part of the metric can be understood as a ``squashed'' 
$3$-sphere. 
In the case that $B=C$, the so-called \emph{biaxial} case, the system enjoys an additional
$U(1)$ symmetry. If $B=C=0$ we have $SO(4)$ symmetry and the unique solution
to the Einstein vacuum equations
is five-dimensional Schwarzschild, 
which we will here refer to as the Schwarzschild-Tangherlini solution.

From the Einstein equations $(\ref{vacuum})$ we derive the following equations:
\begin{equation} \label{eom1}
\partial_u \left( \Omega^{-2} \partial_u r \right) =
-\frac{2r}{3\Omega^2}\left((B_{,u})^2+B_{,u}C_{,u}+(C_{,u})^2\right),
\end{equation}
\begin{equation} \label{eom2}
\partial_v \left( \Omega^{-2} \partial_v r \right) = 
-\frac{2r}{3\Omega^2}\left((B_{,v})^2+B_{,v}C_{,v}+(C_{,v})^2\right),
\end{equation}
\begin{equation} \label{eom3}
-2 \partial_u \partial_v \log \Omega -\frac{3}{r}r_{,uv} = 
B_{,v}\left(2B_{,u}+C_{,u}\right) + C_{,v}\left(2C_{,u}+B_{,u}\right) ,
\end{equation}
\begin{equation} \label{eom4}
\partial_u \partial_v \log \Omega  + 
\frac{3}{r} r_{,uv} + 3\frac{r_{,u}r_{,v}}{r^2} = 
-\frac{\Omega^2 \rho}{2 r^2}-\frac{1}{2}\left(B_{,v}\left(2B_{,u}+C_{,u}\right) 
+ C_{,v}\left(2C_{,u}+B_{,u}\right)\right),
\end{equation}
where $\rho$ denotes the scalar curvature of the group orbit:
\begin{equation} \label{scalcurv}
\rho = e^{2B+2C}+e^{-2B}+e^{-2C}-\frac{1}{2}
e^{-(4B+4C)}-\frac{1}{2}e^{4B} -\frac{1}{2}e^{4C}.
\end{equation}
From these equations we can derive a system
of nonlinear wave equations for the
four quantities $r, \Omega, B,$ and $C$:
\begin{equation} \label{revol}
r_{,uv} = -\frac{1}{3} \frac{\Omega^2 \rho}{r} - \frac{2
  r_{,u}r_{,v}}{r},
\end{equation}
\begin{equation} \label{omegaevol}
\partial_u \partial_v \log \Omega = \frac{\Omega^2 \rho}{2r^2} + 
\frac{3}{r^2} r_{,u} r_{,v} - \frac{1}{2}
  \left(B_{,v}\left(2B_{,u}+C_{,u}\right) +
  C_{,v}\left(2C_{,u}+B_{,u}\right)\right) ,
\end{equation}
\begin{eqnarray} \label{Bevol}
B_{,uv} &=& -\frac{3}{2}\frac{r_{,u}}{r} B_{,v}-
  \frac{3}{2}\frac{r_{,v}}{r} B_{,u}\\
\nonumber 
 &&\hbox{}+\frac{\Omega^2}{3r^2}
  \left(e^{2B+2C}+e^{-4B-4C}-2e^{-2B}-2e^{4B}+e^{-2C}+e^{4C}\right),
\end{eqnarray}
\begin{eqnarray} \label{Cevol}
C_{,uv} &=& -\frac{3}{2}\frac{r_{,u}}{r} C_{,v}-
  \frac{3}{2}\frac{r_{,v}}{r} C_{,u}\\
  \nonumber
  &&\hbox{}+\frac{\Omega^2}{3r^2}
  \left(e^{2B+2C}+e^{-4B-4C}-2e^{-2C}-2e^{4C}+e^{-2B}+e^{4B}\right).
\end{eqnarray}
Note that the last two equations become identical in the biaxial case.
Equations $(\ref{eom1})$ and $(\ref{eom2})$ are to be thought of as constraints which
are preserved by the evolution of $(\ref{revol})$--$(\ref{Cevol})$.

The system of equations $(\ref{eom1})$--$(\ref{Cevol})$ should be compared to the
equations originally derived in~\cite{Bizon} in $r$, $t$ coordinates.

\section{The initial value problem}
Consider an asymptotically flat triaxial Bianchi IX vacuum initial 
data set\footnote{We leave to the reader the correct
formulation of this notion.} $(\mathcal{S},\bar{g},K)$.
Let $(\mathcal{M},g)$ denote the maximal development of $(\mathcal{S},\bar{g},K)$.
By standard arguments, it follows that $(\mathcal{M},g)$ is triaxial Bianchi IX symmetric
in the sense of the previous section. Moreover, the range of the null coordinates
can be chosen to be bounded, defining i.e.~a conformal embedding of
$\mathcal{Q}$ into a bounded subset
of  ${\mathbb R}^{1+1}$. The two possibilities for the
global structure of the image of such an embedding are 
depicted below:
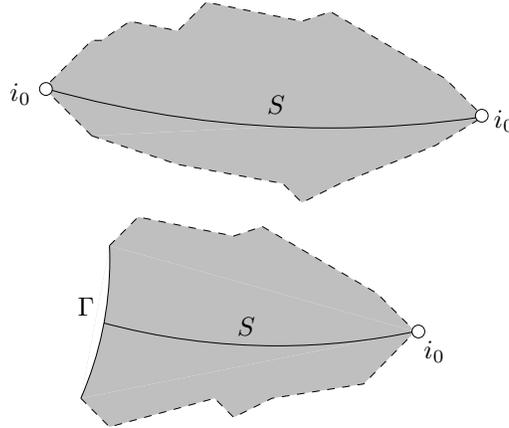
\[
\begin{picture}(0,0)%
\includegraphics{case1.pstex}%
\end{picture}%
\setlength{\unitlength}{3158sp}%
\begingroup\makeatletter\ifx\SetFigFont\undefined%
\gdef\SetFigFont#1#2#3#4#5{%
  \reset@font\fontsize{#1}{#2pt}%
  \fontfamily{#3}\fontseries{#4}\fontshape{#5}%
  \selectfont}%
\fi\endgroup%
\begin{picture}(3801,1599)(3782,-3973)
\put(7583,-3369){\makebox(0,0)[lb]{\smash{\SetFigFont{10}{12.0}{\rmdefault}{\mddefault}{\updefault}{\color[rgb]{0,0,0}$i_0$}%
}}}
\put(5806,-3263){\makebox(0,0)[lb]{\smash{\SetFigFont{10}{12.0}{\rmdefault}{\mddefault}{\updefault}{\color[rgb]{0,0,0}$S$}%
}}}
\put(3782,-3160){\makebox(0,0)[lb]{\smash{\SetFigFont{10}{12.0}{\rmdefault}{\mddefault}{\updefault}{\color[rgb]{0,0,0}$i_0$}%
}}}
\end{picture}
 
\]
\[
\begin{picture}(0,0)%
\includegraphics{case2.pstex}%
\end{picture}%
\setlength{\unitlength}{3158sp}%
\begingroup\makeatletter\ifx\SetFigFont\undefined%
\gdef\SetFigFont#1#2#3#4#5{%
  \reset@font\fontsize{#1}{#2pt}%
  \fontfamily{#3}\fontseries{#4}\fontshape{#5}%
  \selectfont}%
\fi\endgroup%
\begin{picture}(2730,1674)(4854,-4048)
\put(7568,-3496){\makebox(0,0)[lb]{\smash{\SetFigFont{10}{12.0}{\rmdefault}{\mddefault}{\updefault}{\color[rgb]{0,0,0}$i_0$}%
}}}
\put(4854,-3159){\makebox(0,0)[lb]{\smash{\SetFigFont{10}{12.0}{\rmdefault}{\mddefault}{\updefault}{\color[rgb]{0,0,0}$\Gamma$}%
}}}
\put(6106,-3316){\makebox(0,0)[lb]{\smash{\SetFigFont{10}{12.0}{\rmdefault}{\mddefault}{\updefault}{\color[rgb]{0,0,0}$S$}%
}}}
\end{picture}

\]
depending on the number of asymptotically flat ends.
$S$ above denotes $\pi(\mathcal{S})$.
In what follows, the notations $J^+$, closure, etc., will refer to the topology and causal
structure of ${\mathbb R}^{1+1}$.
By the definition of
asymptotic flatness, it follows
that $r$ tends monotonically to infinity along $S$, sufficiently close to the
points labeled $i_0$. Moreover, $\mathcal{Q}\cap J^+(S)$ 
is foliated by constant-$v$ curves
emanating from $S$, and constant-$u$ curves emanating from $S\cup \Gamma$.

\section{Local existence and extension}
We wish to understand those TIPs in $\mathcal{Q}$ 
which do not ``emanate'' from the centre $\Gamma$.
For this, the following local existence theorem in null coordinates shall
suffice for our purposes.
\begin{proposition}
\label{localexistence}
Let $\Omega$, $r$, $B$, and $C$ be functions defined
on $X=[0,d]\times \{0\}\cup \{0\}\times[0,d]$. Let $k\ge0$, and assume 
$r>0$ is $C^{k+2}(u)$
on $[0,d]\times \{0\}$
and $C^{k+2}(v)$ on $\{0\}\times[0,d]$,
assume that $\Omega$, $B$ and $C$ are $C^{k+1}(u)$ on
$[0,d]\times\{0\}$ and $C^{k+1}(v)$
on $\{0\}\times[0,d]$.
Suppose that equations $(\ref{eom1})$, $(\ref{eom2})$
hold initially on $[0,d]\times\{0\}$ and $\{0\}\times[0,d]$, respectively.
Let $|f|_{n,u}$ denote the $C^{n}(u)$ norm of $f$ on
$[0,d]\times \{0\}$, $|f|_{n,v}$ the $C^{n}(v)$ norm of $f$
on $\{0\}\times[0,d]$, etc.
Define
\[
N=\sup\{|\Omega|_{1,u}, |\Omega|_{1,v},
|\Omega^{-1}|_0, |r|_{2,u}, |r|_{2,v}, |r^{-1}|_{0},
|B|_{1,u}, |B|_{1,v}, |C|_{1,u},|C|_{1,v}\}.
\]
Then there exists a $\delta$,
depending only on $N$,
and a $C^{k+2}$ function (unique among $C^2$ functions)
$r$ and $C^{k+1}$ functions (unique among $C^1$ functions)
$\Omega$, $B$, and $C$,
satisfying equations $(\ref{eom1})$--$(\ref{Cevol})$ 
in $[0,\delta^*]\times[0,\delta^*]$, where
$\delta^*=\min\{d,\delta\}$, such that
the restriction of these functions
to $[0,d]\times\{0\}\cup\{0\}\times[0,d]$
is as prescribed.
\end{proposition}
\begin{proof}
The proof is by standard methods and is omitted.
\end{proof}

From the above Proposition and the maximality of the Cauchy development, 
the following extension principle follows.
Given a subset $Y\subset\mathcal{Q}\setminus\Gamma$,
define
\[
N(Y)=\sup\{|\Omega|_{1}, 
|\Omega^{-1}|_0, |r|_{2}, |r^{-1}|_{0},
|B|_{1}, |C|_{1}\},
\]
where, for $f$ defined on $\mathcal{Q}^+$, $|f|_k$ denotes the
restriction of the $C^k$ norm to $Y$.
\begin{proposition}
\label{extensionp1}
Let $p\in\overline{\mathcal{Q}}\setminus\overline{\Gamma}$, and $q\in\mathcal{Q}\cap
I^-(p)$ such that $J^-(p)\cap J^+(q)\setminus \{p\}\subset\mathcal{Q}$, and
$N(J^-(p)\cap J^+(q)\setminus\{p\})<\infty$.
Then $p\in\mathcal{Q}$.
\end{proposition}

\section{The Hawking mass}
A remarkable feature of the system of equations $(\ref{revol})$--$(\ref{Cevol})$
 is the existence
of energy estimates for $B$ and $C$. For this, we first define
the so-called \emph{Hawking mass}
\begin{equation} \label{hawkmass}
m = \frac{r^2}{2} \left(1 + \frac{4 r_{,u} r_{,v}}{\Omega^2} \right).
\end{equation}
We compute the identities:
\begin{equation} \label{hawderu}
\partial_u m = -\frac{4}{3} \frac{r^3}{\Omega^2} r_{,v}
\left[(B_{,u})^2+B_{,u}C_{,u}+(C_{,u})^2\right]+r \cdot r_{,u}
\left[1-\frac{2}{3}\rho\right],
\end{equation}
\begin{equation} \label{hawderv}
\partial_v m = -\frac{4}{3} \frac{r^3}{\Omega^2} r_{,u}
\left[(B_{,v})^2+B_{,v}C_{,v}+(C_{,v})^2\right]+r \cdot r_{,v}
\left[1-\frac{2}{3}\rho\right],
\end{equation}
Note that $\rho$ is bounded above:
\begin{equation} \label{rhobound}
\rho \leq \frac{3}{2}.
\end{equation}
(A straightforward way to show this is to set $x=e^{2B}$, $y=e^{2C}$
and to study the function $\rho(x,y)$. First one shows that 
$\rho(x,y) < \frac{3}{2}$ in the region
\begin{equation}
R = \left\{x \leq \frac{1}{10}, y \leq \frac{1}{10}\right\} \cup \{x \geq 10, y
\geq 10\}.
\end{equation}
Next one determines the critical points of $\rho(x,y)$. It turns out that
that there is only one extremum at $x=1$, $y=1$, which is shown to be
a maximum. This proves $\rho(x,y) \leq \frac{3}{2}$ with equality 
only for the round sphere, $B=C=0$.)

By $(\ref{rhobound})$, we now see that all terms in 
square brackets are manifestly non-negative.
Thus, if, say $\partial_u r<0$ and $\partial_vr \ge0$, we have
\begin{equation}
\label{mmono}
\partial_um\le 0, \partial_vm\ge 0.
\end{equation}

The monotonicity $(\ref{mmono})$ can be compared with the monotonicity
in the $r$-direction for the $(r,t)$ coordinates given by 
formula $(8)$ of~\cite{Bizon}.
 
\section{The regions $\mathcal{R}$, $\mathcal{T}$, and $\mathcal{A}$}
Let us define
the \emph{regular region}
\begin{equation}
\mathcal{R} = \{ p \in \mathcal{Q} \ \ \textrm{such that} \ \
\partial_v r > 0, \partial_u r < 0 \}
\end{equation}
the \emph{trapped region}
\begin{equation}
\mathcal{T} = \{ p \in \mathcal{Q} \ \ \textrm{such that} \ \
\partial_v r < 0, \partial_u r < 0 \}
\end{equation}
and the \emph{marginally trapped region}
\begin{equation}
\mathcal{A} = \{ p \in \mathcal{Q} \ \ \textrm{such that} \ \
\partial_v r = 0, \partial_u r < 0 \}.
\end{equation}
The reader is warned that the term \emph{regular} is meant
with reference to the asymptotically flat end in the direction 
of which the vector $\partial_v$ points. By the results of
the previous section, the inequalities $(\ref{mmono})$ hold
in $\mathcal{R}\cup\mathcal{A}$.
In the next section, we will show how this leads to a stronger extension theorem
than Proposition~\ref{extensionp1}.

\section{Extension in the non-trapped region}
The monotonicity $(\ref{mmono})$ indicates that our system $(\ref{eom1})$--$(\ref{Cevol})$
shares a formal similarity with  spherically symmetric
$3+1$-dimensional Einstein-matter systems, 
for suitable matter fields satisfying the 
dominant energy condition (See~\cite{Mihali1,Mann}).
In particular, one might conjecture that
an extension principle analogous
to the one formulated in~\cite{Mihali1} holds in the non-trapped region.
This is what we show in this section.

We have
\begin{proposition}
\label{extensionp2}
Let $p\in\overline{\mathcal{R}}\setminus\overline{\Gamma}$, and $q\in\mathcal{R}\cup\mathcal{A}
\cap
I^-(p)$ such that $J^-(p)\cap J^+(q)\setminus \{p\}\subset\mathcal{R}\cup\mathcal{A}$. 
Then $p\in\mathcal{R}\cup\mathcal{A}$.
\end{proposition}
\begin{proof}
The proof adapts techniques introduced in~\cite{Mihali2}.
Let us introduce the following notation:
\begin{align}
\nu &= \partial_u r, \\
\lambda &= \partial_v r, \\
\kappa &= -\frac{1}{4} \frac{\Omega^2}{\nu}, \\
\mu &= \frac{2m}{r^2}, \\
\zeta_{B} &= r^{\frac{3}{2}} \partial_u B, \\
\zeta_{C} &= r^{\frac{3}{2}} \partial_u C, \\
\theta_{B} &= r^{\frac{3}{2}} \partial_v B, \\
\theta_{C} &= r^{\frac{3}{2}} \partial_v C.
\end{align}
Note that $\kappa(1-\mu)=\lambda$.

Let $X$ denote $(J^+(q)\setminus I^+(q))\cap\mathcal{Q}$.
Setting $p=(u_1,v_1)$, $q=(u_\epsilon,v_\epsilon)$, we have
$X=\{u_\epsilon\}\times[v_\epsilon,v_1]\cup[u_\epsilon,u_1]\times\{v_\epsilon\}$.
Since $X$ is compact, the quantities
\begin{equation} \label{quantities}
r, \kappa, \lambda, \nu, m, B, C, \zeta_{B}, \zeta_{C}, \theta_{B},
\theta_{C}, \partial_u \Omega, \partial_v \Omega, \partial_v \lambda,
\partial_u \nu
\end{equation}
are uniformly bounded above and below on $X$:
\begin{align}
\label{similar}
\begin{split}
0 < r_0 \leq r \leq R, \\
0 \leq \lambda \leq \Lambda, \\
0 > \nu_0 \geq \nu \geq N, \\
\vert B \vert \leq P_B, \\
\vert C \vert \leq P_C, \\
\vert \theta_{B} \vert \leq T_B, \\
\vert \theta_{C} \vert \leq T_C, \\
\vert \zeta_{B} \vert \leq Z_B, \\
\vert \zeta_{C} \vert \leq Z_C, \\
\vert m \vert \leq M, \\
0 <\kappa_0\le \kappa \leq K, \\
\vert \partial_u \Omega \vert \leq H, \\
\vert \partial_v \Omega \vert \leq H, \\
\vert \partial_u \nu \vert \leq H, \\
\vert \partial_v \lambda \vert \leq H.
\end{split}
\end{align}
By Proposition~\ref{extensionp1}, in view also of the fact that
$\Omega^2=-4\kappa\nu$, to prove Proposition~\ref{extensionp2}, 
it suffices to show
that the quantities
(\ref{quantities}) are uniformly bounded everywhere in $[u_\epsilon,u_1]\times[v_\epsilon,v_1]
\setminus \{ (u_1,v_1)\}$, with bounds similar to $(\ref{similar})$.

We first derive a bound for $r$. Integrating $\nu$ along $u$,
and $\lambda$ along $v$, we obtain
from  $(\ref{similar})$, in view of the signs
of $\nu$, $\lambda$ in $\mathcal{R}\cup\mathcal{A}$, that
\begin{equation}
\label{rbounds}
0<r_0\le r\le R
\end{equation}
in $[u_\epsilon,u_1]\times[v_\epsilon,v_1]
\setminus \{ (u_1,v_1)\}$. A similar argument can
be given for the mass: Integrating (\ref{hawderu}) along $u$ yields
\begin{equation}
m(u^{\star}, v^{\star})-m(u_{\epsilon}, v^{\star}) \leq 0
\end{equation}
and integrating (\ref{hawderv}) along $v$ yields
\begin{equation}
m(u^{\star}, v^{\star})-m(u^{\star},v_{\epsilon}) \geq 0.
\end{equation}
We conclude the bound
\begin{equation}
\label{conclude}
-M \leq m \leq M
\end{equation}
on $[u_\epsilon,u_1]\times[v_\epsilon,v_1]
\setminus \{ (u_1,v_1)\}$.

A bound on $\kappa$ can be derived as follows: 
Note that $\kappa>0$ by definition, in view of the $\nu<0$.
On the other hand, we compute  from $(\ref{eom1})$
\begin{equation}
\label{kappae3is}
\kappa_{,u} = -\frac{1}{6} \frac{r}{\nu^2}{\Omega^2} 
\left((B_{,u})^2+B_{,u}C_{,u}+(C_{,u})^2\right)\le 0.
\end{equation}
Thus, integrating in $u$ from $X$, in view of $(\ref{similar})$, we obtain
\begin{equation}
\label{asteraki}
0<\kappa\le K
\end{equation}
in $[u_\epsilon,u_1]\times[v_\epsilon,v_1]
\setminus \{ (u_1,v_1)\}$.

Next we bound the quantity $\nu$ using the evolution equation
(\ref{revol}), written:
\begin{equation}
\partial_v \nu=
r_{,uv} = -\frac{1}{3} \frac{\Omega^2 \rho}{r} - \frac{2 \nu
  \lambda}{r} = \nu \left( \frac{4 \kappa \rho}{3r} - 
\frac{2\lambda}{r}\right).
\end{equation}
Integrating this equation in $v$ we get
\begin{equation} \label{nueq}
\nu(u^\star, v^\star) = \nu (u^\star, v_\epsilon) \exp \left( \int_{v_\epsilon}^{v^\star} \left(\frac{4 \kappa \rho}{3r} - \frac{2
  \lambda}{r} \right) (u^\star, v) dv \right).
\end{equation}
Since $\rho\le\frac{3}{2}$,
$\lambda\ge 0$, we obtain the upper bound
\begin{equation}
-\nu \leq \vert N \vert \cdot \exp \left(\frac{2 \epsilon K}{r_0} \right) \equiv N^\prime.
\end{equation}
From the above and $(\ref{asteraki})$ it follows that the
quantity $\Omega^2=-4 \kappa \nu$ is also bounded from above. 

To estimate $B$ and $C$, we
revisit the equations (\ref{hawderu}) and (\ref{hawderv}), in view of (\ref{conclude}),
 to infer a-priori integral estimates for derivatives of these quantities. 
 Equation
(\ref{hawderv}) gives
\begin{equation}
\label{massbnd}
\int_{v_{\epsilon}}^{v^{\star}} \left(-\frac{4}{3} \nu
\frac{r^3}{\Omega^2}\left((B_{,v})^2+B_{,v}C_{,v}+(C_{,v})^2\right) +
 \lambda r \left(1-\frac{2}{3}\rho \right) \right)(u^\star, v) \ dv \leq 2M
\end{equation}
and therefore,
since
\[
B_v^2+B_vC_v+C_v^2\ge \frac12B_v^2+\frac12C_v^2\ge 0,
\]
we have
\begin{equation} \label{ap1}
\int_{v_{\epsilon}}^{v^{\star}} \frac{1}{3}
\frac{r^3}{\kappa}(B_{,v})^2 \ (u^{\star},v) \ dv =
\int_{v_{\epsilon}}^{v^{\star}} \frac{1}{3 \kappa}
\left(\theta_{B}\right)^2 \ (u^{\star},v) \ dv \leq 4M.
\end{equation}
Obviously, the same inequality holds with $B$ replaced by $C$. In the same way,
integrating equation (\ref{hawderu}) along $u$ using the
mass-bound (\ref{conclude}) leads to the estimate
\begin{equation} \label{ap2}
\int_{u_{\epsilon}}^{u^{\star}} \frac{1}{3} \left(1-\mu \right)
\left(\frac{\zeta_{B}}{\nu}\right)^2 (-\nu) \ (u, v^{\star}) \ du \leq 4M.
\end{equation}
Again, the same inequality holds with $B$ replaced by $C$.

We may now integrate the equation $B_{,v} = r^{-\frac{3}{2}}{\theta_{B}}$
in $v$ to obtain
\begin{equation}
\begin{split}
\vert B(u^{\star},v^{\star}) \vert \leq \vert B(u^{\star},v_{\epsilon}) \vert
+ \left| \int_{v_{\epsilon}}^{v^{\star}} \frac{\theta_{B}}{r^{\frac{3}{2}}} \left(u^{\star},v
\right) dv \right| \\
\leq P_B + \sqrt{\int \frac{\theta_{B}^2}{\kappa} dv}
\sqrt{\int \frac{\kappa}{r^3} dv} \leq P_B + \sqrt{12M} \sqrt{\frac{K
    \epsilon}{r_0^3}} \equiv P_b,
\end{split}
\end{equation}
where we used the Schwarz inequality in the step from the first to
the second line and (\ref{ap1}) for the last step. 
In a completely analogous fashion--integrating
$C_{,v} = r^{-\frac{3}{2}}\theta_{C}$ in $v$--we obtain the
same bound for $C$. Having bounded $B$ and $C$, it follows 
from (\ref{scalcurv}) that $\rho$ is also bounded in 
$[u_\epsilon,u_1]\times[v_\epsilon,v_1]
\setminus \{ (u_1,v_1)\}$. This enables us to bound $\lambda$. 
Rewriting the evolution equation (\ref{revol}) for $r_{,uv}$ in
terms of quantities we already control we obtain
\begin{equation}
\partial_u\lambda=r_{,uv} = \nu \left( \frac{4 \kappa}{3r} \rho - \frac{2}{r} \kappa
\left(1-\frac{2m}{r^2} \right) \right)
\end{equation}
which we can integrate along $u$. Because we already control all the
quantities appearing in the integrand we immediately obtain a bound 
for $\lambda$ in $[u_\epsilon,u_1]\times[v_\epsilon,v_1]
\setminus \{ (u_1,v_1)\}$:
\begin{equation}
\lambda \leq L.
\end{equation}
The determination of a suitable constant $L$ is left to the reader. 

We turn to bound $|\nu|$ and $\kappa$ from below, away from zero. 
In view of the bound on
$|\rho|$, we may derive immediately from $(\ref{nueq})$ a bound
\[
\nu \le \tilde{\nu}_0<0.
\]
For $\kappa$,
we integrate $(\ref{kappae3is})$, rewritten as
\begin{equation}
\label{ikiyildiz}
\partial_u\kappa=\kappa\left(\frac23r\nu^{-1}(B_u^2+B_uC_u+C_u^2)\right)
\end{equation}
to obtain
\begin{eqnarray*}
\kappa(u,v) &=& \kappa(u_\epsilon,v)\exp{\int_{u_\epsilon}^u\frac23r \nu^{-1}
(B_u^2+B_uC_u+C_u^2)du}		\\
	    &\ge & \tilde{\kappa}_0,
\end{eqnarray*}
where, for the last inequality, we use $(\ref{similar})$ and the bounds proved
above, in particular, the $u$-analogue of $(\ref{massbnd})$.

Finally, we note at this stage that from
(\ref{revol}), it follows
immediately $r_{,uv}$ is bounded in $[u_\epsilon,u_1]\times[v_\epsilon,v_1]
\setminus \{ (u_1,v_1)\}$.

We turn now to bound the derivatives of $B$ and $C$. First let us consider
$\partial_v B$, $\partial_v C$: 
Differentiating $\theta_{B}=r^{\frac{3}{2}} \partial_v B$ in $u$ and using the
evolution equation (\ref{Bevol}) we get
\begin{equation}
\partial_u \theta_{B} = -\frac{3}{2} \frac{\lambda \zeta_{B}}{r} + \frac{\Omega^2}{3\sqrt{r}}
  \left(e^{2B+2C}+e^{-4B-4C}-2e^{-2B}-2e^{4B}+e^{-2C}+e^{4C}\right),
\end{equation}
which can be integrated in $u$ to give
\begin{equation}
\begin{split}
\vert \theta_{B}(u^{\star}, v^{\star}) \vert \leq \vert \theta_{B}(u_{\epsilon},
v^{\star}) \vert + \frac{3}{2} \left| \int_{u_{\epsilon}}^{u^{\star}}
\frac{\lambda \zeta_{B}}{r} \ (u, v^{\star}) \ du \right| + \\
\left| \int_{u_{\epsilon}}^{u^{\star}} \frac{\Omega^2}{3\sqrt{r}}
  \left(e^{2B+2C}+e^{-4B-4C}-2e^{-2B}-2e^{4B}+e^{-2C}+e^{4C}\right) \
  (u, v^{\star}) \ du  \right|. 
\end{split}
\end{equation}
The term in the last line is bounded because we control all quantities
in the integrand. We estimate it say by the constant $F$. For the
second term we use the Schwarz inequality and the a-priori bound
(\ref{ap2}):
\begin{equation}
\begin{split}
\vert \theta_{B} (u^{\star}, v^{\star}) \vert \leq T_B + F + \frac{3}{2} \left| \int_{u_{\epsilon}}^{u^{\star}}
\frac{\nu \kappa (1-\mu) \zeta_{B}}{r \nu} \ (u, v^{\star}) \ du \right| \\
\leq T_B + F + \frac{3}{2} \sqrt{\int_{u_{\epsilon}}^{u^{\star}}
(-\nu) (1-\mu) \left(\frac{\zeta_{B}}{\nu}\right)^2 \ (u, v^{\star}) \
  du} \ \ \sqrt{\int_{u_{\epsilon}}^{u^{\star}}
\frac{(-\nu) \kappa^2 (1-\mu)}{r^2} \ (u, v^{\star}) \ du} \\
\leq T_B + F + \frac{3}{2} \sqrt{12M} \cdot K \cdot \sqrt{r_0^{-1}+Mr_0^{-3}} \equiv V.
\end{split}
\end{equation}
Hence we bounded $\theta_{B}$ and therefore $\partial_v B$. The bound for
$\partial_v C$ is obtained completely analogously. 

Next we turn to $\partial_uB$, $\partial_uC$. Differentiating $\zeta_{B} =
r^{\frac{3}{2}} \partial_u B$ with respect to $v$ using the evolution
equation (\ref{Bevol}), we obtain
\begin{equation}
\partial_v \zeta_{B} = -\frac{3}{2} \frac{\nu \theta_{B}}{r} +\frac{\Omega^2}{3\sqrt{r}}
  \left(e^{2B+2C}+e^{-4B-4C}-2e^{-2B}-2e^{4B}+e^{-2C}+e^{4C}\right).
\end{equation}
Integration in $v$ now yields a bound for $\zeta_{B}$ since all the
quantities on the right have already been shown to be bounded. (Alternatively we
could use the Schwarz inequality and the a-priori bound (\ref{ap1}).)
The bound for $\zeta_{C}$ and therefore $C_{,u}$ is obtained in a completely
analogous manner. Having bounded $B, C$ and their first
derivatives, equation (\ref{Bevol}) yields that $B_{,uv}$ is also
bounded.

Bounds for $\Omega_u$ and $\Omega_v$ follow by integrating
(\ref{omegaevol}) in $v$ and $u$ respectively.
Finally, bounds for $r_{,uu}$ and  $r_{,vv}$ follow from
(\ref{eom1}) respectively (\ref{eom2}) and the previous bounds.

As remarked at the beginning, the proof now follows by applying Proposition~\ref{extensionp1}.
\end{proof}

\section{Null infinity}
Let $\tilde{S}$ be as in the statement of Theorem~\ref{introthe}. 
Without loss of generality, let  the asymptotically
 flat end in question be such that $\partial_v$ points
``outwards''.
We define a set $\mathcal{I}^+\subset 
(\overline{\mathcal{Q}}\setminus\mathcal{Q})\cap J^+(\tilde S)$,
as follows: Let
\begin{equation}
\mathcal{U} = \left\{ u \ : \ \sup_{v \ : \ (u,v) \in \mathcal{Q}^+} r(u,v)
= \infty \right\}.
\end{equation}
For each $u \in \mathcal{U}$, there is a unique $v^\star(u)$ such that
\begin{equation}
(u, v^\star(u)) \in (\overline{\mathcal{Q}} \setminus \mathcal{Q})\cap 
J^+(\tilde S).
\end{equation}
Let the end in question have limit point on $S$ given by $i_0=(\hat{u},V)$.
Then the
\emph{null-infinity} corresponding to $i_0$ is defined as the set
\begin{equation}
\mathcal{I}^+ = \bigcup_{u \in \mathcal{U}:v^\star(u)=V} (u,v^\star(u)).
\end{equation}
Standard arguments show that $\mathcal{I}^+$ is non-empty for the data
considered here.
It is straightforward to show, adapting
\cite{Mihali1}, that $\mathcal{I}^+$ is then a connected ingoing
null-ray with past-limit point $i_0$.  Denote the future
limit point of $\mathcal{I}^+$ by $i^+$. 
A priori, 
it could be that $i^+\in\mathcal{I}^+$.

Adapting~\cite{Mihali1}, one shows from $(\ref{mmono})$ that
the 
\emph{Bondi mass}
\[
M(u)=\limsup_{v\to V} M(u,v)
\]
is a finite (not necessarily continuous) function 
on $\mathcal{I}^+$, non-increasing in $u$. 
We define $M_f=\inf M(u)$ to be the \emph{final Bondi mass}.

\section{Proof of Theorem~\ref{introthe}}
This proof is an adaptation of methods introduced in~\cite{Mihali1}.

As above, let $\tilde{S}$ be as in the statement of Theorem~\ref{introthe}, let $\partial_v$
be the outward direction, and
consider the set 
\[
\mathcal{D}=J^+(\tilde{S})\cap J^-(\mathcal{I}^+)\cap\mathcal{Q}.
\]
This set is non-empty. On the other hand,
by the Raychaudhuri equations $(\ref{eom1})$--$(\ref{eom2})$, and
the assumption that $\partial_ur<0$ along $\tilde{S}$, 
it follows that $\partial_ur<0$ along future-directed constant-$v$ curves in $\mathcal{Q}$
emanating from $\tilde S\cap\{v\ge v(p)\}$,
and
\[
\mathcal{D}\subset\mathcal{R}.
\]
Since by assumption $p\in\mathcal{T}\cup\mathcal{A}$, it follows that
$p\not\in\mathcal{D}$, and thus $\mathcal{D}$ has a non-empty future boundary
in $\mathcal{Q}$.
Denote this boundary $\mathcal{H}^+$. 
Note also that $m\ge  r^2(p)>0$ in $\mathcal{D}$, and thus in particular, $M_f>0$.

Proposition~\ref{extensionp2} shows immediately that $\mathcal{H}^+$ cannot 
terminate before
reaching $i^+$, i.e.,
the Penrose diagram is as:
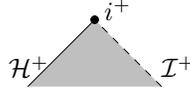
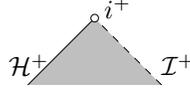
\[
\begin{picture}(0,0)%
\includegraphics{2.pstex}%
\end{picture}%
\setlength{\unitlength}{3158sp}%
\begingroup\makeatletter\ifx\SetFigFont\undefined%
\gdef\SetFigFont#1#2#3#4#5{%
  \reset@font\fontsize{#1}{#2pt}%
  \fontfamily{#3}\fontseries{#4}\fontshape{#5}%
  \selectfont}%
\fi\endgroup%
\begin{picture}(1212,669)(3151,-3448)
\put(3901,-2911){\makebox(0,0)[lb]{\smash{\SetFigFont{10}{12.0}{\rmdefault}{\mddefault}{\updefault}{\color[rgb]{0,0,0}$i^+$}%
}}}
\put(4351,-3361){\makebox(0,0)[lb]{\smash{\SetFigFont{10}{12.0}{\rmdefault}{\mddefault}{\updefault}{\color[rgb]{0,0,0}$\mathcal{I}^+$}%
}}}
\put(3151,-3361){\makebox(0,0)[lb]{\smash{\SetFigFont{10}{12.0}{\rmdefault}{\mddefault}{\updefault}{\color[rgb]{0,0,0}$\mathcal{H}^+$}%
}}}
\end{picture}

\]
or
\[
\begin{picture}(0,0)%
\includegraphics{1.pstex}%
\end{picture}%
\setlength{\unitlength}{3158sp}%
\begingroup\makeatletter\ifx\SetFigFont\undefined%
\gdef\SetFigFont#1#2#3#4#5{%
  \reset@font\fontsize{#1}{#2pt}%
  \fontfamily{#3}\fontseries{#4}\fontshape{#5}%
  \selectfont}%
\fi\endgroup%
\begin{picture}(1212,669)(3151,-3448)
\put(3901,-2911){\makebox(0,0)[lb]{\smash{\SetFigFont{10}{12.0}{\rmdefault}{\mddefault}{\updefault}{\color[rgb]{0,0,0}$i^+$}%
}}}
\put(4351,-3361){\makebox(0,0)[lb]{\smash{\SetFigFont{10}{12.0}{\rmdefault}{\mddefault}{\updefault}{\color[rgb]{0,0,0}$\mathcal{I}^+$}%
}}}
\put(3151,-3361){\makebox(0,0)[lb]{\smash{\SetFigFont{10}{12.0}{\rmdefault}{\mddefault}{\updefault}{\color[rgb]{0,0,0}$\mathcal{H}^+$}%
}}}
\end{picture}

\]
We will first show that the latter is the case, 
i.e.~$i^+\not\in \mathcal{I}^+$, in fact, that the Penrose inequality
\begin{equation}
\label{Penroseevent}
r^2 \leq 2M_f,
\end{equation}
holds
on the event horizon $\mathcal{H}^+$.

To show $(\ref{Penroseevent})$ on $\mathcal{H}^+$, 
one assumes the contrary, i.e.~the existence of a point
$(\widetilde{U}, \widetilde{V})$ with $r^2(\widetilde{U},
\widetilde{V})=R^2 > 2M_f$  on the horizon, and as in~\cite{Mihali1},
one infers (using monotonicity properties of $r$ and $m$, together
with Proposition~\ref{extensionp2}) 
the existence of a neighbourhood of the horizon which is part of the
  regular region: 
\begin{equation}
\mathcal N:=\left[u_0, u^{\prime \prime}\right]\times[\widetilde{V},V) \subset
  \mathcal{R}
\end{equation}
with $u_0 < \widetilde{U} < u^{\prime \prime}$. In particular this
neighbourhood can be chosen such that there exists 
an $R^\prime < R$ with the property
 that in $\left[\widetilde{U}, u^{\prime \prime}\right]\times[\widetilde{V},V) \subset
  \mathcal{R}$
\begin{equation} \label{estnbh}
r \geq R^\prime \ \ \ \textrm{and} \ \ \ 1-\frac{2m}{r^2} \geq
1-\frac{2M}{\left(R^\prime\right)^2}
\end{equation}
holds.
The last step is to show that 
for any $u^{\star} \in \left[u_0, u^{\prime \prime}\right]$,
$\lim_{v^\star \rightarrow \infty} r(u^\star, v^\star) = \infty
$, i.e. $\mathcal{H}^+$ cannot be the event horizon, as defined,
a contradiction.

To show this last step, having shown $(\ref{estnbh})$, we proceed as follows:
Integrating
  (\ref{hawderu}) along $u$ from $u_0$ to a point $u^\star <
  u^{\prime \prime}$ we obtain the estimate
\begin{equation}
\sup_{\overline{v} \geq \widetilde{V}} \int_{u_0}^{u^{\star}}
  \frac{4r^3}{3 \Omega^2} \lambda
  \left((B_{,u})^2+B_{,u}C_{,u}+(C_{,u})^2\right) (\overline{u},
  \overline{v}) \ d\overline{u} \leq M
\end{equation}
which can be written as
\begin{equation}
\sup_{\overline{v} \geq \widetilde{V}} \int_{u_0}^{u^{\star}}
  \frac{r^3}{3 (-\nu)} (1-\mu)
  \left((B_{,u})^2+B_{,u}C_{,u}+(C_{,u})^2\right) (\overline{u},
  \overline{v}) \ d\overline{u} \leq M.
\end{equation}
Taking (\ref{estnbh}) into account we can derive the estimate
\begin{equation} \label{estfin}
\sup_{\overline{v} \geq \widetilde{V}} \int_{u_0}^{u^\star}
\frac{1}{3} \frac{r
  \left((B_{,u})^2+B_{,u}C_{,u}+(C_{,u})^2\right)}{\nu}
(\overline{u}, \overline{v}) d\overline{u} \geq \frac{-M}{(R^\prime)^2 -2M}
\end{equation}
valid for any $u^\star \in [u_0, u^{\prime \prime})$.
Integrating (\ref{ikiyildiz}),
we obtain
\begin{equation}
\kappa(u^\star, v^\star) \geq
\kappa(u_0, v^\star) \cdot \exp
\left(\frac{-2M}{(R^\prime)^2-2M} \right)
\end{equation}
and therefore
\begin{equation} \label{fh}
\lambda (u^\star, v^\star) \geq \left(1-\frac{2M}{R^2} \right) \exp
\left(\frac{-2M}{(R^\prime)^2-2M} \right) \ \lambda (u_0, v^\star).
\end{equation}
Integrating (\ref{fh}) in $v$, we see that $\lim_{v^\star
  \rightarrow V} r(u^\star, v^\star) \rightarrow \infty$, since $\lim_{v^\star
  \rightarrow V} r(u_0, v^\star) \rightarrow \infty$ on the right by
  the definition of $\mathcal{I^+}$. We conclude $(u^\star, V) \in
  \mathcal{I}^+$. Therefore,  $\mathcal{H}^+$ is not the event horizon
  and we have arrived at the desired contradiction.

The only thing left in the proof of Theorem~\ref{introthe} is 
to show the
the completeness of $\mathcal{I}^+$.
(Completeness here refers to an adaptation in~\cite{Mihali1} of the concept
defined in~\cite{Christodoulou}.)
We have to show that the suitable normalized affine
length, as measured from a fixed outgoing null curve $u=u_0$,
of the ingoing null-curves $v=const$ in $J^-(\mathcal{I}^+)$
tends to infinity as $v \rightarrow V$. 
More precisely, we define the vector field
\begin{equation}
X(u,v) = \frac{\Omega^2(u_0, v)}{\Omega^2(u,v)}
\frac{\partial}{\partial u}
\end{equation}
on $J^-(\mathcal{I}^+) \cap \mathcal{Q}^+$. Note that this vector
field is parallel along all ingoing null-rays and along the curve
$u=u_0$. We will show 
\begin{equation}
\lim_{v \rightarrow V} \int_{u_0}^{\widetilde{U}} \left(X(u,v)
\cdot u \right)^{-1} du = \infty.
\end{equation}

From equation (\ref{eom1}) we can derive
\begin{eqnarray} \label{he1}
\Omega^2(u,v)\Omega^{-2}(u_0,v) &=& \nu(u,v)\nu^{-1}(u_0,v)\cdot\\
\nonumber
&&\hbox{}\cdot\exp \left(\int_{u_0}^u
\frac{2r}{3\nu}
\left((B_{,u})^2+B_{,u}C_{,u}+(C_{,u})^2\right) (\overline{u},v)
d\overline{u}\right).
\end{eqnarray}
Let $M$ be the Bondi-mass at $u_0$. We choose an $R$ such that
$R^2 > 2M \geq 2M_f$ and consider the curve $\{r=R\} \cap
J^-(\mathcal{I}^+)$. For sufficiently large $v_0 < V$, all ingoing
null-curves with $v> v_0$ intersect $\{r=R\} \cap
J^-(\mathcal{I}^+)$ at a unique point $(u^\star(v),v)$, depending on
$v$. 

Analogously to (\ref{estfin}) we derive the bound
\begin{equation} \label{he2}
\int_{u_0}^u
\frac{2r}{3\nu}
\left((B_{,u})^2+B_{,u}C_{,u}+(C_{,u})^2\right) (\overline{u},v)
d\overline{u} \geq \frac{-2M}{R^2-2M} \, ,
\end{equation}
which we use to estimate
\begin{align}
\begin{split}
\int_{u_0}^{\widetilde{U}} &\left(X(u,v) \cdot u \right)^{-1} du 
\geq \int_{u_0}^{u^\star(v)}  \left(X(u,v) \cdot u \right)^{-1} du \\
&=\nu^{-1}(u_0,v) \int_{u_0}^{u^\star(v)} \exp 
\left(\int_{u_0}^u \frac{2r}{3(\partial_u r)}
\left((B_{,u})^2+B_{,u}C_{,u}+(C_{,u})^2\right) (\overline{u},v)
d\overline{u}\right) \nu du \\
&\geq \frac{r(u_0,v)-R}{(-\nu)(u_0,v)} \exp
\left(\frac{-2M}{R^2-2M} \right).
\end{split}
\end{align}

Since $r(u_0,v) \rightarrow \infty$ as $v \rightarrow \infty$ we only
need to show that $(-\nu)(u_0,v)$ is uniformly bounded in $v$.
The quantity
\begin{equation}
\frac{\nu}{1-\mu}
\end{equation}
satisfies
\begin{equation}
\partial_v \frac{\nu}{1-\mu} = \frac{
  \nu}{1-\mu} \frac{2r}{3\lambda
  }\left((B_{,v})^2+B_{,v}C_{,v}+(C_{,v})^2\right)
\end{equation} 
which integrates to
\begin{equation} \label{cb}
\begin{split}
\frac{-\nu}{1-\mu}(u_0,v) = \\
 \exp \left(
\int_{v_0}^v \frac{2r}{3} \frac{1}{\lambda}
\left((B_{,v})^2+B_{,v}C_{,v}+(C_{,v})^2\right)
(u_0,\overline{v}) d\overline{v}\right)
\frac{-\nu}{1-\mu}(u_0,v_0).
\end{split}
\end{equation}
We can choose $v_0$ (so large) such that
\begin{equation}
1-\frac{2M}{(r(u_0,v_0))^2} > 0.
\end{equation}
Set $R^\prime = r(u_0,v_0)$. Analogously to (\ref{estfin}) and
(\ref{he2}) we derive the bound
\begin{equation}
\int_{v_0}^v \frac{2r}{3} \frac{1}{\lambda}\left((B_{,v})^2+B_{,v}C_{,v}+(C_{,v})^2\right)
(u_0,\overline{v}) d\overline{v} \leq \frac{2M}{(R^\prime)^2-2M}
\end{equation}
which enables us to obtain from (\ref{cb}) the estimate
\begin{equation}
-\nu (u_0,v) \leq \left(1-\frac{2M}{(R^\prime)^2} \right)^{-1} \exp
 \left( \frac{2M}{ (R^\prime)^2-2M} \right)
\end{equation}
for $v \geq v_0$, which in turn shows uniform boundedness of $(-\nu)(u_0, v)$ 
in $v$.

We have shown above the future completeness of $\mathcal{I}^+$. 
The past completeness follows by
standard results assuming sufficiently tame initial asymptotic behaviour. In particular,
it is immediate in the case that $B$, $C$, $\nabla B$, $\nabla C$ vanish initially
outside a compact set.

\section{Proof of Corollary~\ref{stab}}
Let $S$ denote the projection of an arbitrary
spherically symmetric Cauchy surface in Schwarzschild, and
let $\tilde{S}$ denote the projection of a second asymptotically flat spherically symmetric
Cauchy surface, with the property
that $\tilde{S}$ contains a $p$ satisfying the conditions of Theorem~\ref{introthe}. 
(Such Cauchy surfaces clearly exist.) By Cauchy stability,
sufficiently small triaxial Bianchi IX perturbations of Schwarzschild data on 
$\pi^{-1}(S)$ yield
solutions $(\mathcal{M}',g')$
possessing a triaxial Bianchi IX symmetric
Cauchy surface $\tilde{S}'$ with geometry arbitrarily close to that of
 $\tilde{S}$, in particular, also satisfying the assumptions
of Theorem~\ref{introthe}. We apply thus this theorem.

Finally, we note that the Hawking mass on $\tilde{S}'$ is arbitrarily close
to the constant value $M$ it takes on Schwarzschild, i.e.~we have
$M-\epsilon\le m\le M+\epsilon$ on $\tilde{S}$.
By the monotonicity $(\ref{mmono})$, it follows that this bound is preserved
in $J^+(\tilde{S}')\cap J^-(\mathcal{I}^+)$. It is this statement--together
with the stability of the Penrose diagram and the completeness of $\mathcal{I}^+$--that 
we term ``orbital stability''.

\section{Final comments}
Besides orbital stability, one is interested in what could be
called \emph{asymptotic stability} of the Schwarzschild family, i.e.~the statement
that perturbations of a Schwarzschild initial data set asymptotically
approach another Schwarzschild solution. An even more ambitious problem would be to
understand the rates of approach, as in~\cite{Igor}. These
problems remain open.

Another interesting and partly related
question is to understand the structure of the outermost apparent horizon.\footnote{Here,
outermost is with respect to the double null foliation.}
In analogy to~\cite{Mihali1}, we may define this as the set
\begin{equation}
\mathcal{A}^\prime = \{(u,v) \in \mathcal{A} \ : \ (u^\star, v) \in
\mathcal{R} \ \textrm{for all} \ u^\star < u \ \textrm{and} \ \exists
u^\prime \ : \ (u^\prime,v) \in J^-(\mathcal{I}^+) \cap \mathcal{Q}\cap J^+(\tilde{S}) \}
\end{equation}
As in~\cite{Mihali1}, $\mathcal{A}'$ is now easily shown
to be an achronal curve intersecting all 
ingoing null curves for $v \geq v_0$ for sufficiently large $v_0$.
In addition, one shows easily that on
$\mathcal{A}^\prime$, the Penrose inequality $(\ref{Penroseevent})$ holds.
There are many other issues, however, which are not settled:
Is it a connected set in a neighborhood of $i^+$? Is it ``generically'' a strictly
spacelike curve in a neighborhood of $i^+$? 
Does it terminate at $i^+$ in the topology of the Penrose diagram? 
For more on these questions, the reader can consult the literature
on so-called \emph{dynamical horizons}, in particular~\cite{Ashtekar}.

\section{Acknowledgments}
The problem addressed in this paper was posed in
 a talk~\cite{Gibbons} of Gary Gibbons at the Newton
Institute of the University of Cambridge. 
G.H.~thanks Gary Gibbons for helpful discussions. M.D.~thanks 
Piotr Bizon and Bernd Schmidt.

\end{document}